# Influence of $Yb^{3+}$ on the structural, electrical and optical properties of sol-gel synthesized Ni-Zn nanoferrites

**N. Jahan[1], M. N. I. Khan[2], F. -U. -Z. Chowdhury[1], A. K. M. Akhter Hossain[3], S. M. Hoque[2], M. A. Matin[4], M. N. Hossain[4], M. M. Hossain[1] and M. M. Uddin[1*]**

[1]Department of Physics, Chittagong University of Engineering and Technology (CUET), Chattogram-4349, Bangladesh
[2]Materials Science Division, Atomic Energy Center, Dhaka 1000, Bangladesh.
[3]Department of Physics, Bangladesh University of Engineering and Technology (BUET), Dhaka-1000, Bangladesh.
[4]Department of Glass and Ceramic Engineering, Bangladesh University of Engineering and Technology (BUET), Dhaka-1000, Bangladesh.

## ABSTRACT

Polycrystalline Yb-substituted Ni-Zn nanoferrites with the compositions of $Ni_{0.5}Zn_{0.5}Yb_xFe_{2-x}O_4$ ($x$ = 0.00, 0.04, 0.08, 0.12, 0.16 and 0.20) have been synthesized using sol-gel auto combustion technique. Single phase cubic spinel structure has been confirmed by the X-ray diffraction (XRD) patterns. Larger lattice constants of the compositions are found with increasing $Yb^{3+}$ concentration while the average grain size (52–18 nm) has noticeable decrease as $Yb^{3+}$ content is increased. The presence of all existing elements as well as the purity of the samples has also been confirmed from energy dispersive X-ray spectroscopic (EDS) analysis. Frequency dependent dielectric constant, dielectric loss, dielectric relaxation time, AC and DC resistivity of the compositions have also been examined at room temperature. The DC resistivity value is found in the order of $10^{10}$ Ω-cm which is at least four orders greater than the ferrites prepared by conventional method. This larger value of resistivity attributes due to very small grain size and successfully explained using the Verwey and deBoer hopping conduction model. The contribution of grain and grain boundary resistance has been elucidated using Nyquist/Cole-Cole plot. The study of temperature dependent DC resistivity confirms the semiconducting nature of all titled compositions wherein bandgap (optical) increases from 2.73 eV to 3.25 eV with the increase of Yb content. The high value of resistivity is of notable achievement for the compositions that make them a potential candidate for implication in the high frequency applications where reduction of eddy current loss is highly required.

* Corresponding author: mohi@cuet.ac.bd (M.M. Uddin)

## 1. Introduction

Spinel ferrites have unique and versatile properties that are very attractive to the researchers. The prime attention is put on the innovation of novel materials with new and low-cost synthesis techniques that have enriched the properties of the materials and fit them with a new technological demand [1]. Nanocrystalline magnetic materials are one of the extensive researches pertaining to their application in the field of biomedical, technological, industrial, defense applications, etc. [2]. Due to their noteworthy physical and chemical properties, nanoferrites are vastly used in high density magnetic storage, electronic and microwave gadget, sensors, magnetically conducted drug delivery [3-5], magneto-caloric applications [6], catalysis [7], hyperthermia [8, 9], biocompatible magnetic nanoparticles [10,11], bioactive molecule separation [12] and magneto-optical devices [13]. Amongst various types of ferrites low cost soft polycrystalline Ni-Zn ferrite is demandable due to its high frequency applications, high dielectric constant, low dielectric loss, high resistivity, high Curie temperature, large magnetic permeability, mechanical strength and chemical stability at relatively low frequencies [14]. The electrical and magnetic properties of nanoferrites can be mediated by substituting electrical insulators rare-earth (RE) materials with high electrical resistivity [15]. In addition the properties of ferrites were significantly altered due to the chemical composition, preparation techniques, sintering time and temperature, cation distribution on tetrahedral and octahedral sites and types of doping impurity and levels [16, 17] etc. Many researchers have already been doped RE ions in the Ni-Zn ferrites systems using various preparation techniques. Ghafoor *et. al.* substituted $Ho^{3+}$ with Ni-Zn ferrite synthesized using the conventional ceramic method with the composition of $Ni_{0.7}Zn_{0.3}Ho_{2x}Fe_{2-2x}O_4$ [18] and studied both electrical and magnetic properties. Beside various elements are substituted in the Ni-Zn ferrites system such as Gd [19], Tb [1], La, Yb, Dy, Ce [15], Nd [20], Gd, Nd, Yb, Lu [21], Y, Eu, Gd [22], Er [23], Y [14], Sm [24], Pr [25]. Among them, some researchers have studied both magnetic and electrical, and some of them elucidated only structural and electrical properties of RE substituted Ni-Zn ferrites. It is noteworthy that the RE ions with larger ionic radius generally occupy the octahedral site in the spinel structure with limited solubility. These larger ions distort the structure and induce strain and consequently improve the electrical and magnetic properties [26-29]. To predict the feasibility of the Yb-substituted Ni-Zn nanoferrites for the practical applications and establishing a database as well, a systematic study is necessary. To the best of our knowledge, the physical properties of Yb-substituted Ni-Zn nanoferrites synthesized by sol-gel auto combustion method in the form of $Ni_{0.5}Zn_{0.5}Yb_xFe_{2-x}O_4$ ($x$ = 0.00, 0.04, 0.08, 0.12, 0.16 and 0.20) have not been reported. Therefore, we are intended to uncover the effect of RE ions $Yb^{3+}$ substitution for the $Fe^{3+}$ ions in the $Ni_{0.5}Zn_{0.5}Yb_xFe_{2-x}O_4$ ($x$ = 0.00, 0.04, 0.08, 0.12, 0.16 and 0.20) ferrites for the first time. The electrical, dielectric and optical along with physical properties in detail for the sol-gel auto combustion derived Yb-substituted Ni-Zn nanoferrites at room temperature have been presented in the following sections.

## 2. Experimental

### 2.1 Synthesis route

Nanocrystalline Ni-Zn ferrite with the composition $Ni_{0.5}Zn_{0.5}Yb_xFe_{2-x}O_4$ ($x$= 0.00, 0.04, 0.08, 0.12, 0.16 and 0.20) was synthesized by sol-gel auto combustion technique. To prepare the samples, analytical grade nitrate salts were taken as raw materials. The precursor salts were weighted according to the stoichiometric ratio. Precursor salts were homogeneously dissolved in small amount of ethanol. Then all materials were stirred by a magnetic stirrer until they are dissolved homogeneously. The solution was then heated at 80°C in a magnetic heater until a viscous gel was formed. After the formation of gel, it was dried in a low temperature oven at temperature 250°C for 5 hrs. Thereafter the dry ash was milled by an agate mortar and pestle. 5% polyvinyl alcohol solution was then mixed with the calcined powder as a binder and samples of desired shapes (pellet and ring) were prepared by applying 10 kN pressure using hydraulic press. Finally, the samples were sintered at 700°C with step of 5°C/min for 5 hrs in air and cooled naturally.

### 2.2 Measurements

The purity of phase and the structural parameters of the ferrites were analyzed by X-ray diffractometer (XRD) (Rigaku Smart Lab with Cu-$K_\alpha$ radiation ($\lambda$= 1.5406 Å) at room temperature (RT). The scanning was done in the range between 15° to 70°; the voltage and current were maintained at 40mV and 40mA, respectively. A high resolution Field Emission Scanning Electron Microscopy (FESEM) (JEOL JSM-7600F) was used to take the micrographs with the EDS. Dielectric measurements were done by a Wayne Kerr precision impedance analyzer (6500B) in the frequency range of 10-120 MHz with a drive voltage of 0.5 V at RT. DC resistivity at RT was measured by two probe method using 6514 KEITHLEY electrometer.

## 3. Results and discussion

### *3.1. Structural properties*

The XRD pattern of Yb-substituted Ni-Zn ferrites (NZYF)) with the chemical composition of $Ni_{0.5}Zn_{0.5}Yb_xFe_{2-x}O_4$ ($x$= 0.00, 0.04, 0.08, 0.12, 0.16 and 0.20) are shown in Fig. 1 (a). The sharp and well-defined peaks are observed that indicate the single phase spinel structure of all samples. No impurity peaks are identified in all samples. The most intense peaks (311) at around 35.6$^o$ of all samples have been observed and depicted in Fig. 1(b) for better understanding. It is evident that peaks are broadening gradually with increasing the RE ions ($Yb^{3+}$) indicating the successful substitution of the ions in place of $Fe^{3+}$ in the compositions. The prominent (311) peaks broadening are associated with f-orbitals that are deeply concealed instead of d-orbitals (showing off) results the sharp broadening occurs [30]. It is perceived from the Fig. 1(b) that the prominent peak (311) first shifts to the higher 2θ value and afterward it backs to the lower 2θ value. It reveals that the (311) interplaner spacing $d$ initially decreases and then increases with increasing $Yb^{3+}$ contents in the composition.

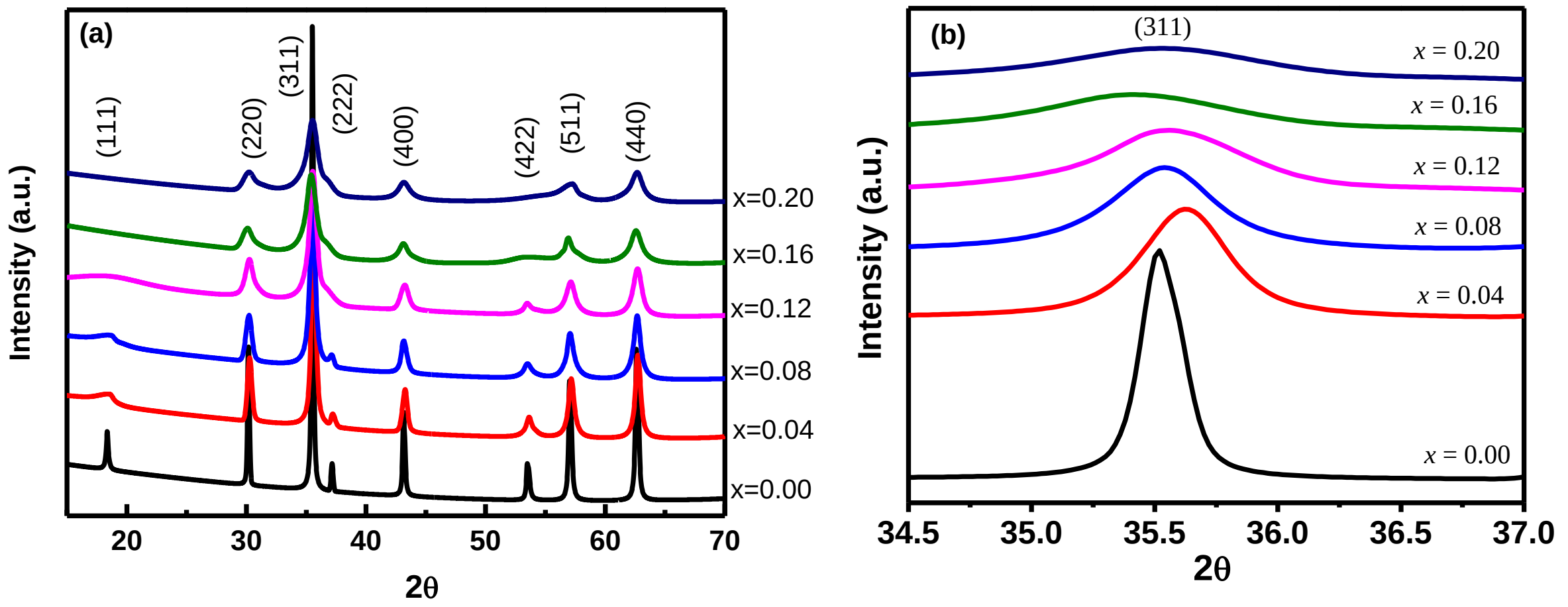

Fig.1: (a) The XRD pattern of $Ni_{0.5}Zn_{0.5}Yb_xFe_{2-x}O_4$ ($x$= 0.00, 0.04, 0.08, 0.12, 0.16 and 0.2). (b) The prominent peaks of 311 plane at around 2θ= 35.6° for various $x$ contents.

The lattice parameter and crystallite size of the compositions have been measured using the XRD spectra. The most intense peaks (311) are used to calculate the average crystallite size of the compositions using the Debye Scherrer's equation, $D = \frac{0.9\lambda}{\beta cos\theta}$, where D is the average crystallite size, λ is the X-ray wavelength of the source (1.5406Å), β is the full width at half maximum (FWHM) and θ is the Bragg's angle [31, 32]. The calculated crystallite sizes are presented in Table 1. The crystallite size decreases from 64 to 11nm with increasing concentration of $Yb^{3+}$ is shown in Fig. 2. The crystalline anisotropy induces that produces strain inside the cell volume during the substitution of $Yb^{3+}$ ions in place of $Fe^{3+}$ due to the difference of ionic radius between the $Yb^{3+}$(0.868 Å) and $Fe^{3+}$ (0.67 Å). The sharp declination in the crystallite size is observed from 64 to 25 nm for $x$ = 0 to 0.04 $Yb^{3+}$ contents thereafter the declination happen to slow to reach lowest size of 11 nm at $x$= 0.20 that makes sense from broadening of the intense peaks of (311) plane shown in Fig. 1(b). It seems that the $Yb^{3+}$ ions decrease the degree of crystallinity and lesser the crystal size of the Ni-Zn ferrites.

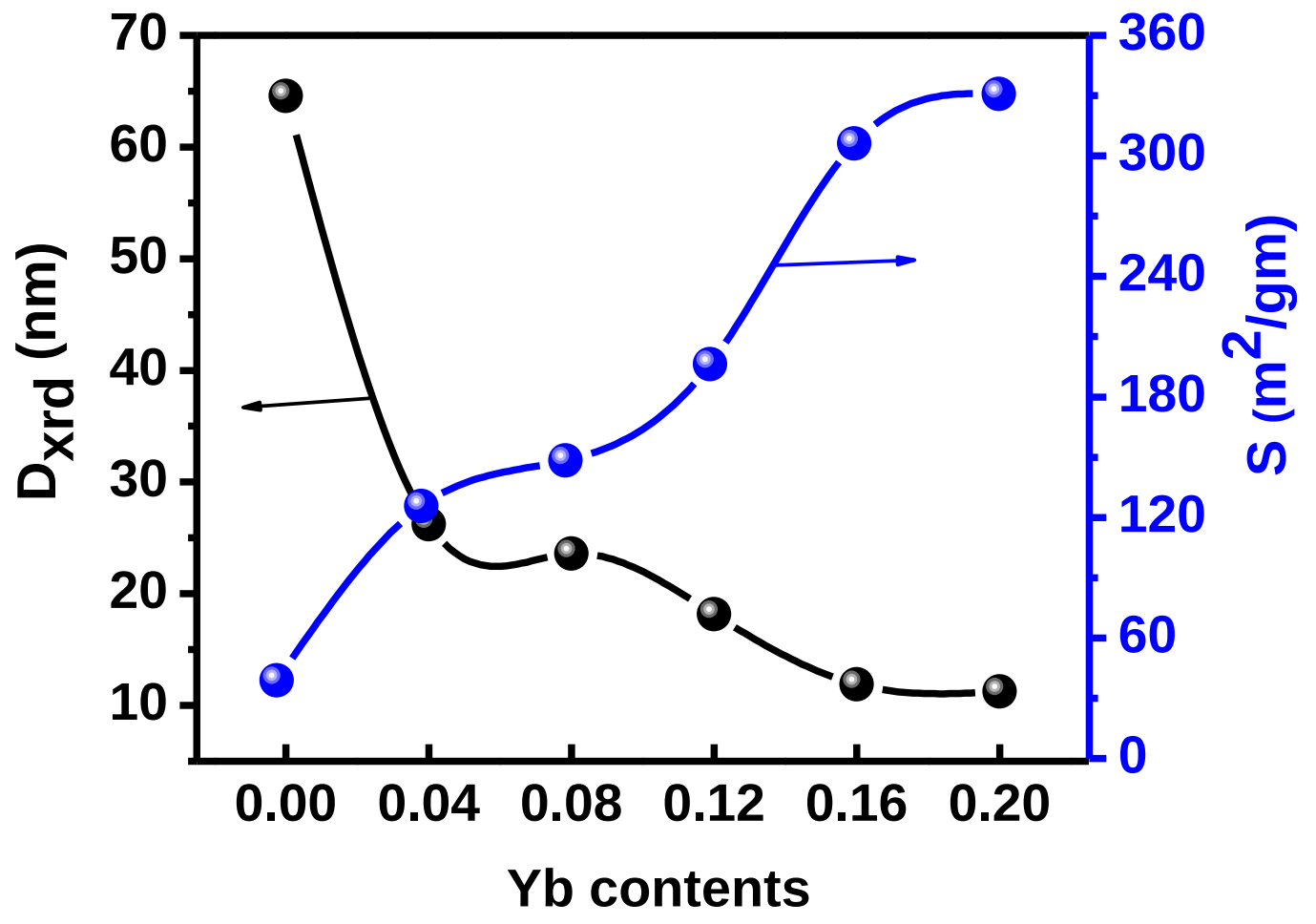

Fig.2: The crystallite size and specific surface area of $Ni_{0.5}Zn_{0.5}Yb_xFe_{2-x}O_4$ ($x$ = 0.00, 0.04, 0.08, 0.12, 0.16 and 0.2) compositions with various Yb concentration.

The specific surface area (*SSA*) provides the information regarding catalytic applications of these ferrites. The *SSA* has been calculated using the equation, $SSA = \frac{6000}{D\rho_b}$, where *D* is the average crystallite size and $\rho_b$ is the bulk density [33]. The *SSA* increases with rises the $Yb^{3+}$ ions content in the composition while the grain sizes decreases and changes are found to be from 39 to 331 $m^2$/gm with grain sizes 64 to 11 nm at $x$= 0.00 to 0.20 contents in the composition, respectively. This is due to fact that the same volume contains more grain due to the smaller size consequently. Moreover, the grain size reduces almost ~17% with increasing $x$ contents and corresponding *SSA* upsurges ~12% which makes sense relation between grain size and the SSA of the compositions.

The lattice constant '$a$' of all samples has been calculated using the relation: $a = d\sqrt{h^2 + k^2 + l^2}$ where $h$, $k$ and $l$ are the Miller indices and d is the interplanar distance of the crystal planes. To evaluate lattice constants, the Nelson-Riley (N-R) extrapolation method has been used. The N-R function, $F(\theta)$, is [34] $F(\theta)=\frac{1}{2}\left[\frac{\cos^2\theta}{\sin\theta}+\frac{\cos^2\theta}{\theta}\right]$. Fig. 3 shows the variation of lattice constants (theoretical and experimental) with the Yb contents. It can be seen that the $a_{exp}$ first lower than that of parent Ni-Zn ferrite (from 8.393 to 8.388 Å) and then increases up to $x$= 0.12 and further decreases till $x$= 0.20 but still higher than parent composition. It is consistent with peak (311) shifts with respect to 2θ shown in Fig. 1b. It can be explained that the RE ions $Yb^{3+}$ do not enter octahedral (B) site at $x$ = 0.04 even they prefer to occupy $B$ sites for replacing $Fe^{3+}$resulting B-site radius ($R_B$) has not changed (Table 1). Therefore the $a_{exp}$ declines at $x$= 0.04 and then increases consequently the $R_B$ has also been increased.

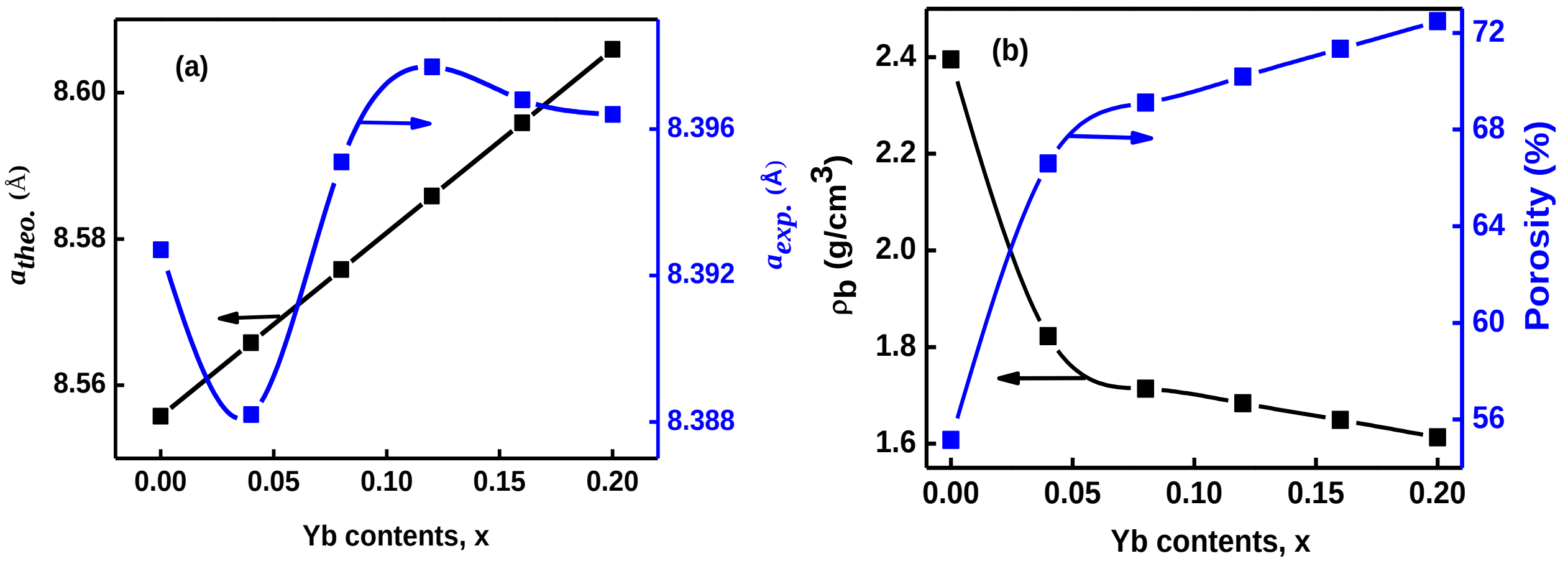

Fig. 3: Variation of (a) lattice constants, $a$ (experimental and theoretical) and (b) bulk density $\rho_b$ and porosity, $P$ of $Ni_{0.5}Zn_{0.5}Yb_xFe_{2-x}O_4$ with Yb contents.

The theoretical lattice constant ($a_{theo}$) for the compositions has been calculated using the equations considering a possible distribution of over $A$- and $B$-sites, $a_{th} = \frac{8}{3\sqrt{3}}\left[(r_A + R_0) + \sqrt{3}(r_B + R_0)\right]$, where $r_A$ and $r_B$ indicate the mean ionic radius of tetrahedral ($A$) and octahedral ($B$) sites, respectively, $R_0$ is the radius of the oxygen ion (1.32 Å). The Ni-Zn is a mixed spinel ferrite where the $Ni^{2+}$ ions prefer octahedral site and $Zn^{2+}$ ions prefer tetrahedral site and $Fe^{3+}$ occupy both tetrahedral and octahedral sites. The values of ionic radius $r_A$ and $r_B$ have been calculated using the relation [35]: $r_A = C_{AFe}r(Fe^{3+})+C_{AZn}r(Zn^{2+})$ and $r_B = 0.5[C_{BFe}r(Fe^{3+}) + C_{BNi}r(Ni^{2+}) + C_{BYb}r(Yb^{3+})]$. The oxygen positional parameter ($u$) has also been calculated using the formula: $u = \left[\frac{1}{a_{th}\sqrt{3}}(r_A + R_0) + \frac{1}{4}\right]$. A fair disagreement between $a_{\text{theo}}$ and $a_{\text{exp}}$ has been observed (Fig. 3a), however the average $a_{\text{exp}}$ value of the composition moderately agrees with the $a_{\text{theo}}$. This can be understood from the following fact that in theoretical calculation, perfect unit cell having cations are in regular arrangement and well-distributed is considered while in the experimental case defects and thermal effects are surely associated therefore an anomaly could be observed.

The following equations [35] are used to calculate the bond lengths of tetrahedral sites ($R_A$), octahedral sites ($R_B$), tetrahedral edge length (R), shared and unshared octahedral edge length R′ and R″, respectively and tabulated in Table 1:

$$R_A = a\sqrt{3}\left(\delta + \frac{1}{8}\right), R_B = a\left(\frac{1}{16} - \frac{\delta}{2} + 3\delta^2\right)^{1/2},$$

$$R = a\sqrt{2}(2u - 0.5), R' = a\sqrt{2}(1 - 2u) \text{and } R'' = a\sqrt{4u^2 - 3u + {}^{11}/_{16}} \text{ [35],}$$

where $\delta = u\text{-}u_{\text{ideal}}$, δ is the inversion parameter that signifies departure from ideal oxygen parameter ($u_{\text{ideal}}$= 0.375 Å) and $a$ is the experimental lattice constant. It is seen from the Table 1 that the average ionic radius of $r_A$ is constant while the average size of $r_B$ increases with Yb concentration, since a larger ionic radius of $Yb^{3+}$ is substituted in place of lower size of $Fe^{3+}$ ions. The tetrahedral bond length ($R_A$) decreases whereas the octahedral bond length ($R_B$) remains same value at $x$= 0.04 and then increases with increasing the $Yb^{3+}$ concentration. The tetrahedral edge length (R) decreases however both the shared and unshared octahedral edge length increase with increasing concentration of $Yb^{3+}$ ions that are in good agreement with Ni substituted Mg-Zn ferrites system [35].

The X-ray density (theoretical density), bulk density and porosity are calculated by using the following equations, respectively: $\rho_x = \frac{8M}{N_A a^3} g/cm^3$ , $\rho_b = \frac{M}{V} g/cm^3$ ,and $P = \left(\frac{\rho_x - \rho_b}{\rho_x}\right) \times 100\%$, where $N_A$ is Avogadro's number ($6.02 \times 10^{23}$ $mol^{-1}$), $M$ is the molecular weight, $V$ (= $\pi r^2 h$) is the volume of the samples, $r$ and $h$ are the radius and height of the samples and are presented in Table 1. It reveals that the X-ray density (bulk density) increases (decreases) with increasing Yb contents in the composition. This increase is due to the dependency of molecular weight and lattice parameter [1]. The $\rho_x$ is inversely proportional to the $a_{\text{exp}}$; therefore increasing trend of $\rho_x$ is very usual. It appears that larger ionic radius $Yb^{3+}$ ions enter into the cell in the place of smaller radius $Fe^{3+}$ ions, which obviously extend the volume of the cell, resulting $\rho_b$ decreases ($\rho_b \propto 1/V$) with increasing Yb contents.

*3.2. Microstructure study*

The electrical and magnetic properties are strongly inspired by the microstructure of ferrites. The morphological study of the composition $Ni_{0.5}Zn_{0.5}Yb_xFe_{2-x}O_4$ (x = 0.00, 0.04, 0.08, 0.12, 0.16 and 0.20) has been performed using the FESEM and shown in Fig. 4. The FESEM images show the homogeneous, spherical and slightly agglomerated grain size [36].

The average grain size of the compositions has been estimated using ImageJ software shown in Table 2. It decreases with increasing of larger ionic radius Yb contents in the compositions. It is suggested that for $Yb^{3+}$ ions, more energy is desired to penetrate into the lattice for the formation of the $Yb^{3+}$ - $O^{2-}$ bond, which yields the $Yb^{3+}$ - $O^{2-}$ bond energy is larger as compared to the $Fe^{3+}$ - $O^{2-}$ bond [33]. Consequently smaller grain size has been observed with increasing $Yb^{3+}$ substitution in the Ni-Zn ferrites. It is consistent with the reported $Pr^{3+}$ RE ions doped Ni-Zn ferrites system **[**25**]**. The $Yb^{3+}$ substituted samples are needed more energy to complete grain crystallization and growth, therefore it is evident that the $Yb^{3+}$ substituted ferrites are more thermally stable than pure Ni-Zn ferrites.

The energy dispersive X-ray spectroscopic (EDS) analysis has been elucidated to endorse the absence of unwanted elements in the studied compositions. The EDS spectra are depicted in Fig. 5. The peak of Fe, Ni, Zn and Yb in the EDS pattern confirms their presence in the composition and the purity of the samples and ensures no unwanted element is being there. The metal cations and anions existing in the compositions have been calculated using EDS spectra and presented in Table 2. The calculated cation-anion ratio of all samples is in fair agreement with reported metal cation-anion ration (3:4) [35].

Table 1: Basic structural parameters of $Ni_{0.5}Zn_{0.5}Yb_xFe_{2-x}O_4$.

| Yb content ($x$) | A-site | B-site | Crystallite size (nm) | $r_A$ (Å) | $r_B$ (Å) | $a_{theo}$ (Å) | $a_{expt.}$ (Å) | $\rho_x$ (g/cm$^3$) | $\rho_b$ (g/cm$^3$) | $P$ (%) | $u$ (Å) | $R_A$ (Å) | $R_B$ (Å) | $R$ (Å) | $R'$ (Å) | $R''$ (Å) |
|---|---|---|---|---|---|---|---|---|---|---|---|---|---|---|---|---|
| 0.00 | $Zn_{0.5}$ $Fe_{0.5}$ | $[Ni_{0.5}Fe_{1.5}]$ $O_4^{2-}$ | 64.58 | 0.745 | 0.697 | 8.556 | 8.393 | 5.34 | 2.395 | 55.157 | 0.3907 | 2.025 | 1.985 | 3.308 | 2.677 | 3.035 |
| 0.04 | $Zn_{0.5}$ $Fe_{0.5}$ | $[Ni_{0.5}Fe_{1.46}Yb_{0.04}]O_4^{2-}$ | 26.20 | 0.745 | 0.701 | 8.566 | 8.388 | 5.46 | 1.822 | 66.600 | 0.3906 | 2.022 | 1.985 | 3.302 | 2.685 | 3.038 |
| 0.08 | $Zn_{0.5}$ $Fe_{0.5}$ | $[Ni_{0.5}Fe_{1.42}Yb_{0.08}]O_4^{2}$ | 23.60 | 0.745 | 0.705 | 8.577 | 8.395 | 5.55 | 1.714 | 69.110 | 0.3905 | 2.021 | 1.988 | 3.300 | 2.692 | 3.042 |
| 0.12 | $Zn_{0.5}$ $Fe_{0.5}$ | $[Ni_{0.5}Fe_{1.38}Yb_{0.12}]O_4^{2-}$ | 18.15 | 0.745 | 0.709 | 8.587 | 8.398 | 5.65 | 1.684 | 70.185 | 0.3904 | 2.019 | 1.990 | 3.297 | 2.700 | 3.045 |
| 0.16 | $Zn_{0.5}$ $Fe_{0.5}$ | $[Ni_{0.5}Fe_{1.34}Yb_{0.16}]O_4^{2}$ | 11.88 | 0.745 | 0.713 | 8.598 | 8.397 | 5.75 | 1.649 | 71.343 | 0.3903 | 2.016 | 1.991 | 3.293 | 2.707 | 3.048 |
| 0.20 | $Zn_{0.5}$ $Fe_{0.5}$ | $[Ni_{0.5}Fe_{1.30}Yb_{0.20}]O_4^{2}$ | 11.24 | 0.745 | 0.717 | 8.608 | 8.396 | 5.86 | 1.613 | 72.480 | 0.3902 | 2.014 | 1.992 | 3.289 | 2.715 | 3.052 |

Table 2: Average grain size and cation-anion ratio of $Ni_{0.5}Zn_{0.5}Yb_xFe_{2-x}O_4$

| Yb contents ($x$) | Average grain size , D (nm) | Cation-anion ratio |
|---|---|---|
| 0.00 | 52.06 | 5.36:1.30 |
| 0.04 | 24.29 | 3.59:3.08 |
| 0.08 | 24.12 | 2.46:4.20 |
| 0.12 | 18.97 | 2.55:4.12 |
| 0.16 | 17.97 | 2.36:4.3 |
| 0.20 | 17.50 | 2.92:3.76 |

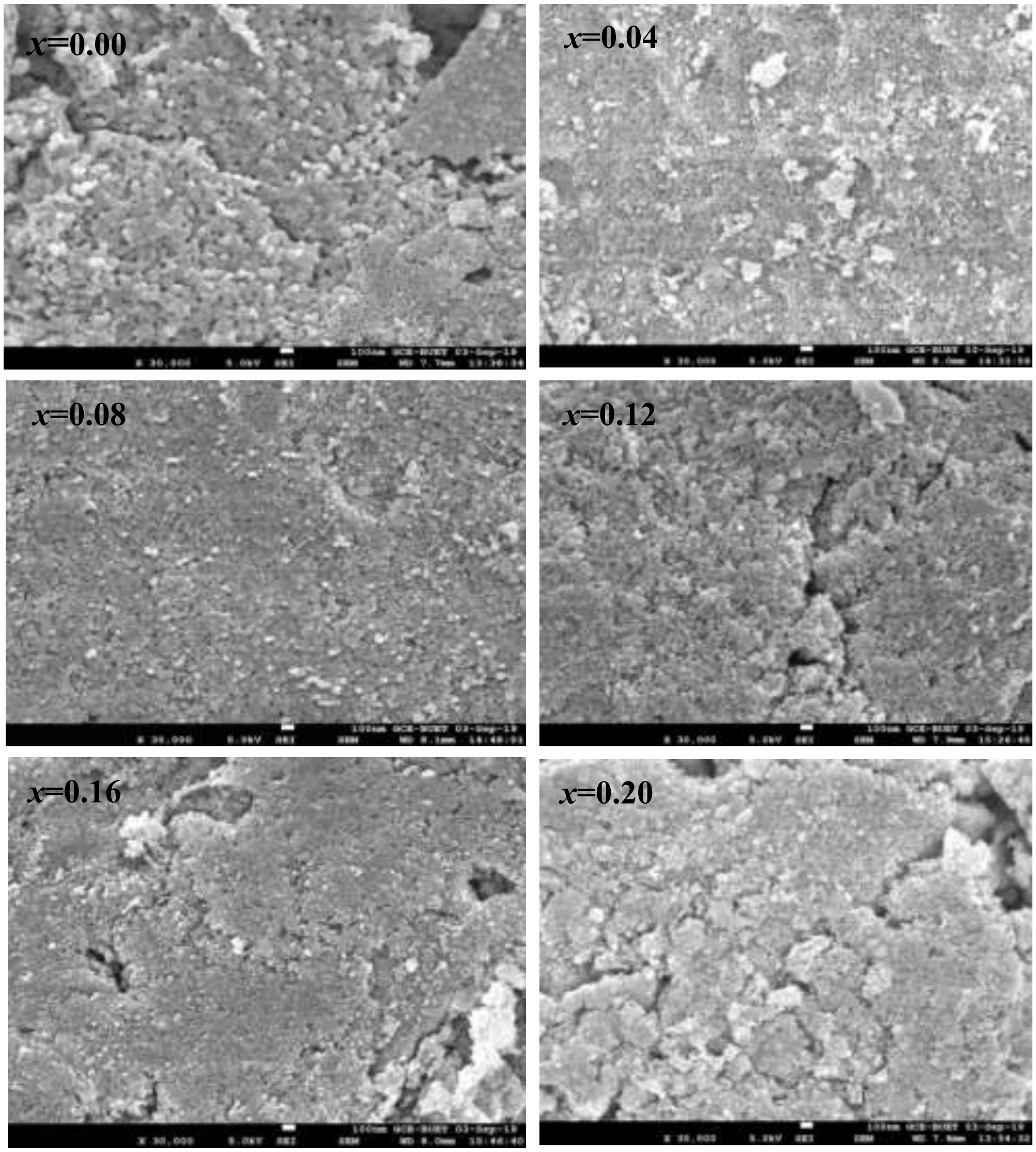

Fig. 4: The FESEM micrographs of $Ni_{0.5}Zn_{0.5}Yb_xFe_{2-x}O_4$ ($x$= 0.00, 0.04, 0.08, 0.12, 0.16 and 0.2) ferrites sintered at 700 °C.

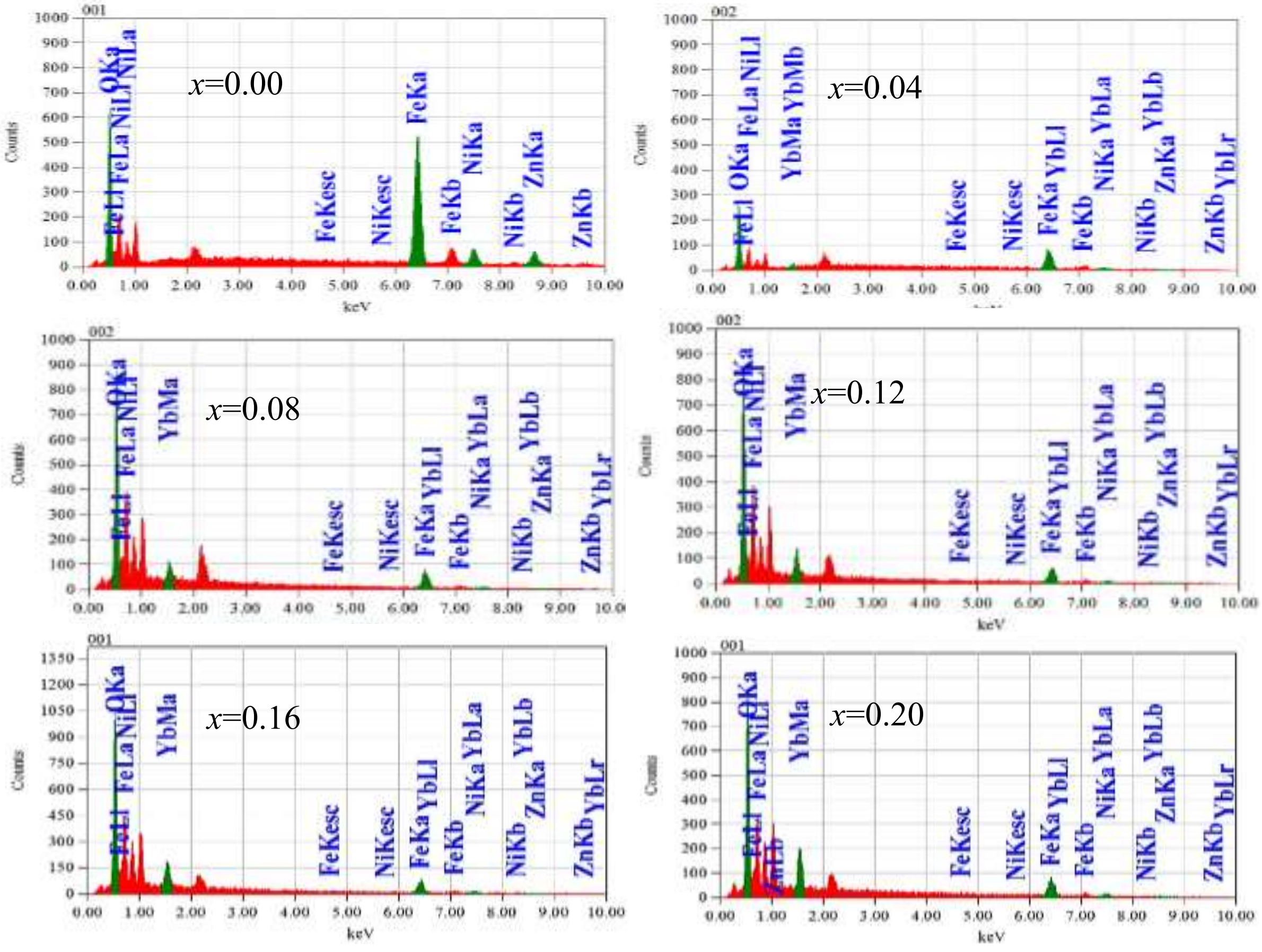

Fig. 5: The EDS pattern of $Ni_{0.5}Zn_{0.5}Yb_xFe_{2-x}O_4$ (x= 0.00, 0.04, 0.08, 0.12, 0.16 and 0.20) ferrites sintered at 700 °C.

The particle size distribution profile of nanoparticles or molecules in a suspension can be estimated using the dynamic light scattering (DLS) technique. The Brownian motion of particles or molecules in suspension is scattered at different intensities by incident of laser light. The velocity of the Brownian motion is determined from the analysis of these intensity fluctuations and consequently the particle size has been measured using the Stokes-Einstein relationship as follows: $D_h = \frac{k_B T}{6\pi\eta D_t}$ , where $D_h$ is the hydrodynamic diameter of the particles, $D_t$ is the translational diffusion coefficient (measured by DLS), $k_B$ is Boltzmann's constant, T is the temperature, η is dynamic viscosity of the suspension [37]. The particle size of the NZYF has been measured by the DLS [Zetasizer (ZEN 3600)] and shown in Fig. 6a along with average grain size ($D_{FESEM}$) of the NZYF measured by the FESEM (Table 2). It is seen that the particle size is higher than that of the $D_{\mathrm{FESEM}}$ that can be explained considering relation between crystallite, grains and particles size in a materials. Crystallite is a single crystal of powder form while grain is single crystal in a bulk/thin film form and a particle is thought as agglomerate which consists of many grains with clear grain boundaries separated each other (inset of Fig. 6b). Therefore, the $D_{\mathrm{xrd}}$ and $D_{\mathrm{FESEM}}$ is almost same in a size that is represented inset of Fig. 6a for the NZYF with different Yb contents. Fig. 6b depicts the number of grains contains in a particle (particle size/ $D_{\mathrm{FESEM}}$) with variation of Yb contents for the NZYF. It is clear that number of grains in a particle increases with increasing Yb content which is consistent with previous discussion (beginning of section 3.2).

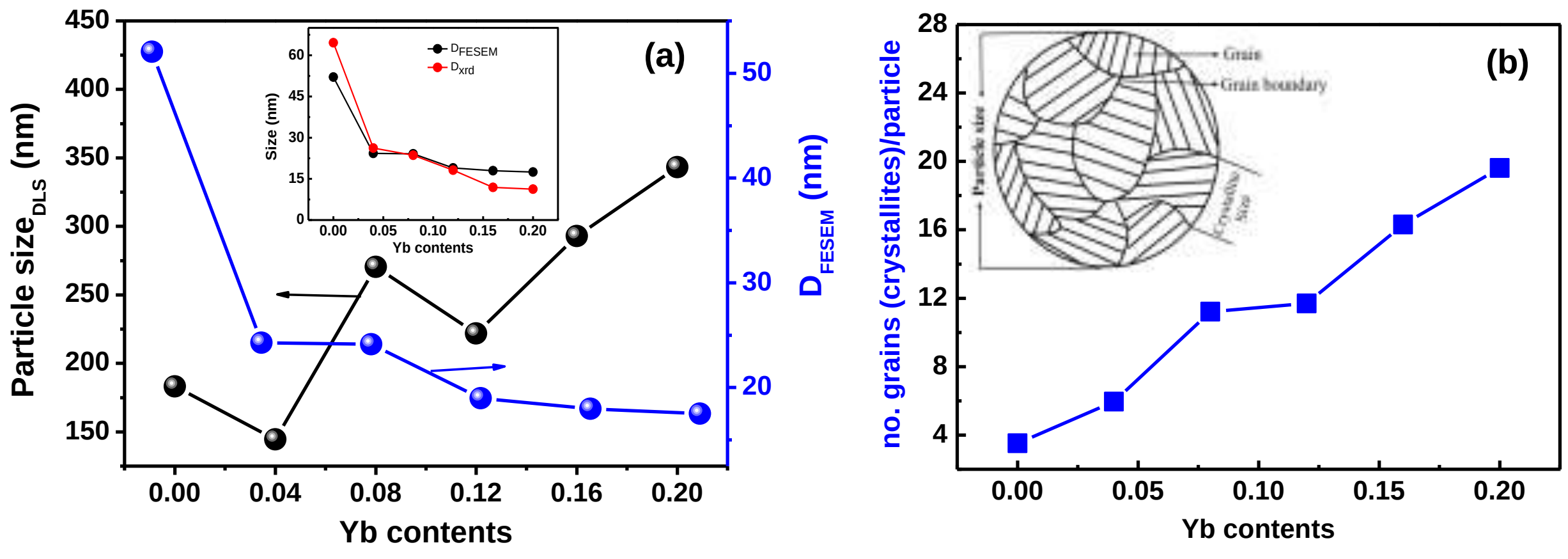

Fig. 6: a) The Yb dependent particle size and average grain size of the NZYF. Inset: the variation of crystallite size ($D_{\mathrm{xrd}}$) and average grain size ($D_{\mathrm{FESEM}}$) as a function of Yb contents. b) Number of grains in a particle as a function of Yb contents. Inset: Schematic of particle, grains and crystallites in nano structured material.

### *3.3. Fourier Transform Infrared (FTIR) spectroscopy*

Fig. 7 shows the FTIR spectra of composition $Ni_{0.5}Zn_{0.5}Yb_xFe_{2-x}O_4$ ($x$= 0.00, 0.04, 0.08, 0.12, 0.16 and 0.20) at room temperature in the range from 350 to 1000 $cm^{-1}$. The peak positions are observed in the expected regions that endorse the spinel structure formed by the sol-gel auto combustion technique. Two different major absorption bands for each sample are noted in the spectra.

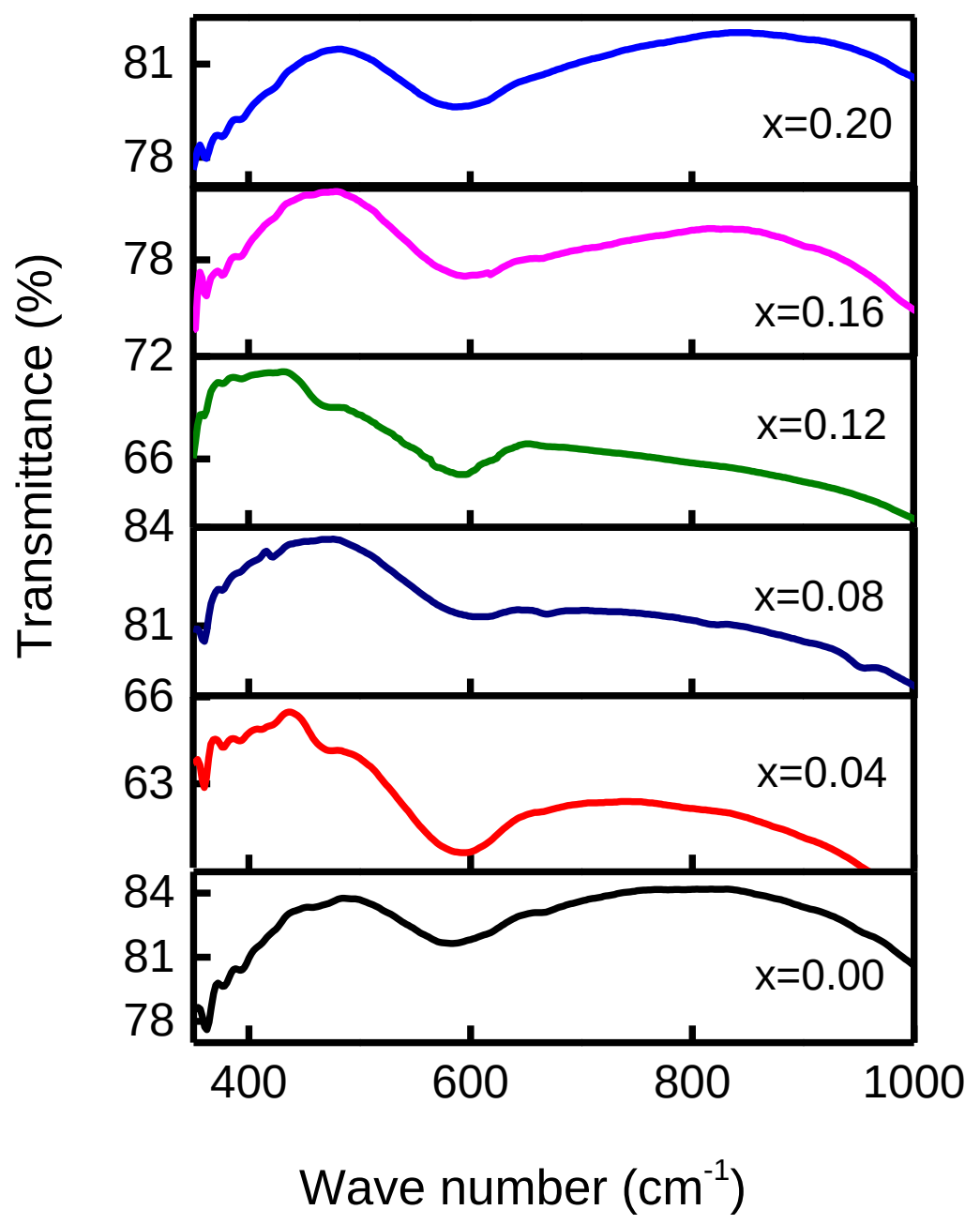

Fig. 7: The FTIR spectra of $Ni_{0.5}Zn_{0.5}Yb_xFe_{2-x}O_4$($x$= 0.00, 0.04, 0.08, 0.12, 0.16 and 0.20).

The band $\nu_1$is constructed by the stretching vibration of the tetrahedral (*A*-sites) M-O bonds and the band $\nu_2$ is constructed in octahedral (*B*-sites) by the M-O vibrations. The bands $\nu_1$ are noted around 595 $cm^{-1}$ correspond to M–O bond in the *A*-sites and $\nu_2$ are noted around 360 $cm^{-1}$ correspond to M–O bond in the *B*-sites, respectively. The high frequency and low frequency absorption bands and their corresponding force constants for all the samples of the series are tabulated in Table 3. The shifting of band frequency is due to the difference of M-O ($Fe^{3+}$-$O^{2-}$) bond. The high frequency absorption band increases with the substituted Yb ions due to the lattice distortion and weakening of the M-O bond. The octahedral band position $\nu_2$ has a tendency to decrease since it is well known that if the site radius increases, the fundamental frequency decreases. Therefore, the center frequency has to shift towards the lower frequency. Similar result has been observed for Ce-substituted Ni-Zn ferrite system [38].The interatomic bonding strength is indicating by the force constant which is defined by $F_c= 4\pi^2c^2\nu^2m$, where $\nu$ is the vibrational frequency, $c$ is the speed of light in free space, and '*m*' is the reduced mass of the metal and oxygen system, $Fe^{3+}$and $O^{2-}$ ions which are equivalent to 2.061 × $10^{-23}$ gm [39]. The successful synthesis of $Ni_{0.5}Zn_{0.5}Yb_xFe_{2-}$ ${}_xO_4$ ($x$= 0.00, 0.04, 0.08, 0.12, 0.16 and 0.20) ferrites has been endorsed by the XRD, FESEM and FTIR).

Table 3**:** Tetrahedral ($\nu_1$) and octahedral ($\nu_2$) band position, tetrahedral and octahedral force constant with average force constant of $Ni_{0.5}Zn_{0.5}Yb_xFe_{2-x}O_4$ ($x$= 0.00, 0.04, 0.08, 0.12, 0.16 and 0.20).

| Yb ($x$) | $\nu_1$ ($cm^{-1}$) | $\nu_2$ ($cm^{-1}$) | $F_cT\times10^5$ (dynes/cm) | $F_cO\times10^5$ (dynes/cm) | $K_{av}\times10^5$ (dynes/cm) |
|---|---|---|---|---|---|
| 0.00 | 581 | 361 | 2.472 | 0.954 | 1.713 |
| 0.04 | 591 | 360 | 2.558 | 0.949 | 1.753 |
| 0.08 | 607 | 360 | 2.698 | 0.949 | 1.824 |
| 0.12 | 593 | 358 | 2.575 | 0.994 | 1.757 |
| 0.16 | 603 | 360 | 2.663 | 0.949 | 1.806 |
| 0.20 | 591 | 360 | 2.558 | 0.949 | 1.753 |

### *3.4. Electrical Properties*

#### *3. 4. 1 DC Resistivity*

The temperature dependent resistivity of the samples sintered at 700 °C has been carried out by two-probe method in the temperature range from 30 °C to 400 °C, shown in Fig. 8. Inset reveals the value of $\rho_{dc}$ at 30 °C for various Yb contents. The measured values are to be order of $10^{10}$ Ω-cm that is greater than at least four order of magnitude compared to the ferrites samples prepared by usual method. Resistivity is extremely depends on the grain size that smaller grain comprises a vast number of grain boundaries those act as barriers to the electron flow. High observed resistivity values in this study are hence attributed to obtain smaller grain size (Table 2) of the compositions synthesized by sol-gel auto combustion technique **[**40**]**. This high value of resistivity is of notable achievement for this composition that makes them a potential candidate for implication in the high frequency applications to reduce eddy current loss.

It is also found that the resistivity declines exponentially with temperature. This indicates the semiconductor nature for all the samples. Due to the thermally generated charge carriers, the resistivity may be decreased. By using the Verwey and deBoer hopping conduction model, the variation of resistivity can be explained [41]. The polaron hopping between $Fe^{3+}$ and $Fe^{2+}$occurs at the *B*-site accordingly the conduction takes place. The electron hopping between *B*- and *A*-sites are very negligible as compared to *B*-site hopping since the distance between two ions at *B*-sites is smaller than the distance at different sites (*A* and *B*) [42]. The charge carriers hopping of *A*-sites are not negligible due to the availability of $Fe^{3+}$ at the tetrahedral site and throughout the process $Fe^{2+}$ ion produced will take the octahedral sites [43].

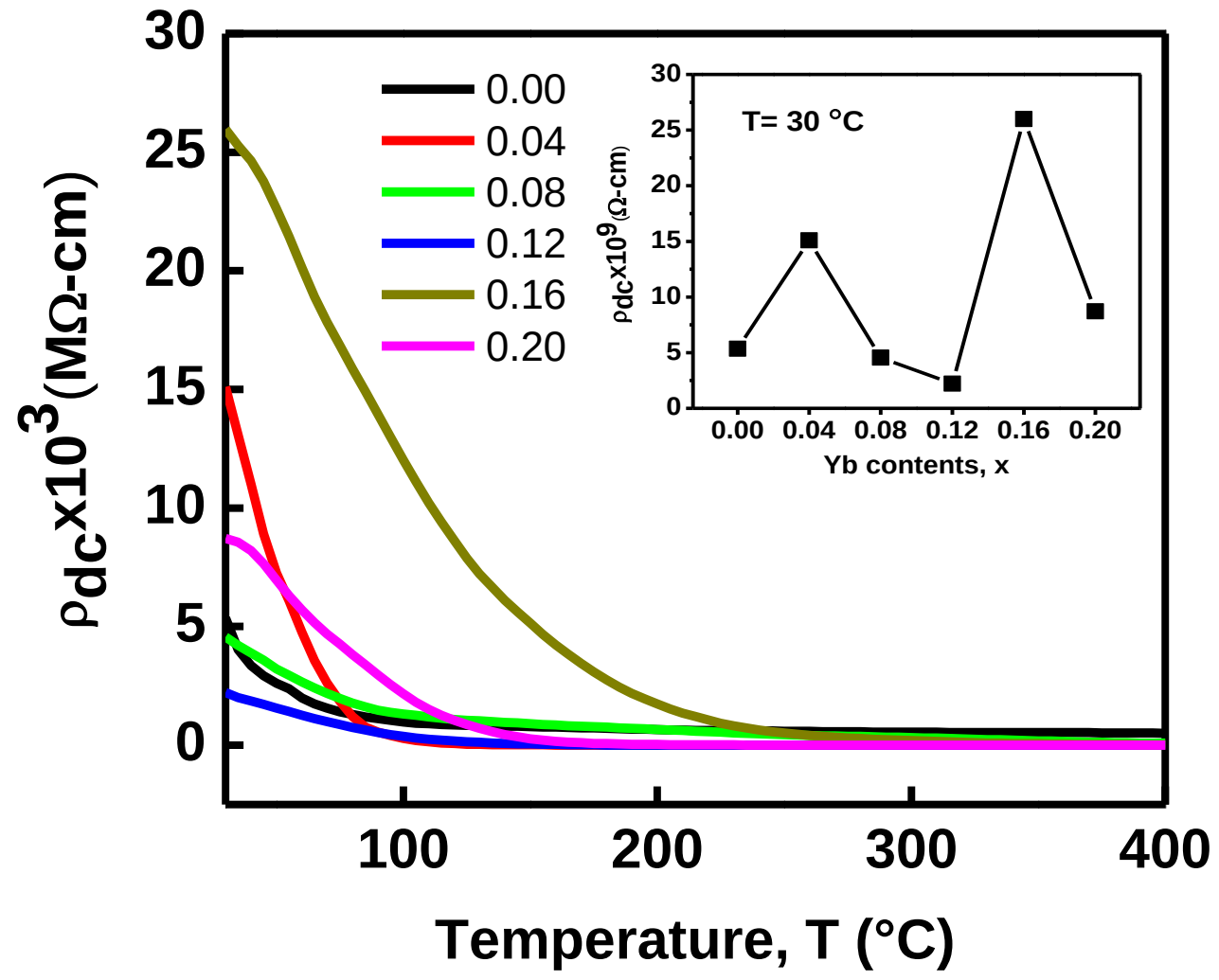

Fig. 8: Temperature dependent DC resistivity of $Ni_{0.5}Zn_{0.5}Yb_xFe_{2-x}O_4$ ($x$= 0.00, 0.04, 0.08, 0.12, 0.16 and 0.20) sintered at 700 °C.Inset shows the Yb content dependent $\rho_{dc}$ value at 30 °C.

*3. 4. 2 AC Conductivity*

Room temperature frequency dependent AC conductivity of $Ni_{0.5}Zn_{0.5}Yb_xFe_{2-x}O_4$ ($x$= 0.00, 0.04, 0.08, 0.12, 0.16 and 0.20) compositions at a fixed frequency 100 Hz has been illustrated in Fig. 9(a).The ac conductivity demonstrations flat at low frequency region, while it illustrates dispersion at high frequency region. Usually, the total conductivity can be articulated by the band and hopping parts [44],

$$\sigma_{total}\ (\omega) = \sigma_{dc} + A\omega^n$$

where the first term is DC conductivity or frequency independent while the second term is frequency dependent and associated with the dielectric relaxation, A is a constant, ω is the angular frequency and the frequency exponent *n* is the dimensionless quantity.

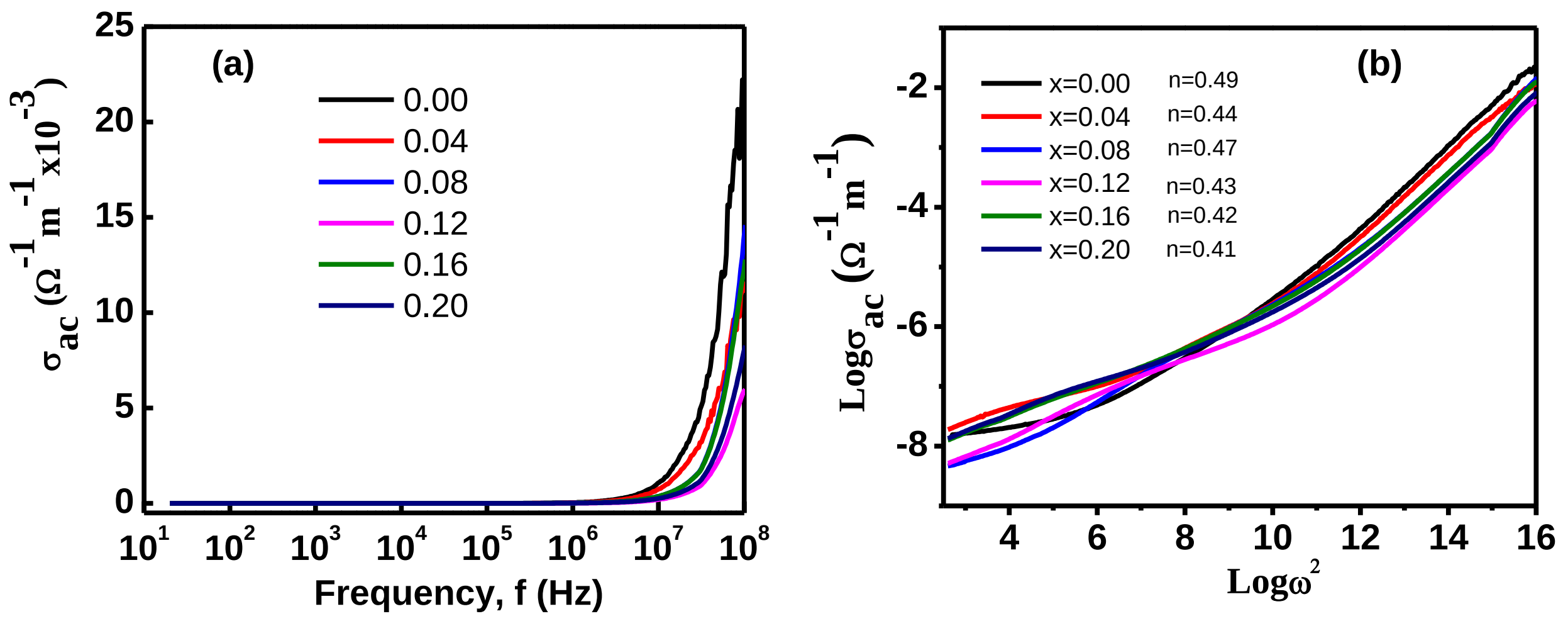

Fig. 9: (a) Room temperature frequency dependence of AC conductivity (b) log $\sigma_{AC}$ vs log $\omega^2$ curve of $Ni_{0.5}Zn_{0.5}Yb_xFe_{2-x}O_4$ ($x$= 0.00, 0.04, 0.08, 0.12, 0.16 and 0.20) sintered at 700 °C.

The conduction mechanism of frequency dependent AC electrical conductivity can be illuminated by the Maxwell-Wagner two layer model and Koop's phenomenological theory. The theory states, the conductive grains are alienated by resistive grain boundaries. At lower frequency, the grain boundaries are more functionalize which leads to low conduction. The grains are predominant at higher frequencies and increase the electron hopping between $Fe^{3+}$ and $Fe^{2+}$ ions hence improves the hopping conduction. The availability of $Fe^{3+}$ and $Fe^{2+}$ ions at octahedral sites is responsible for conduction [45] as well as dielectric polarization. As the substituent ion $Yb^{3+}$ increases, the $Fe^{3+}$ decreases at *B*-sites which reduce the electron exchange between $Fe^{3+}$and $Fe^{2+}$. The plot of log $\sigma_{ac}$ as a function of log $\omega^2$ is illustrated in Fig. 9 (b). The slope of the curves yields the value of *n* that provides information regarding the conduction mechanism of the compositions. The value of *n* indicates i) the conductivity is DC conductivity when n is zero (0), ii) hopping of charge carriers when *n* is in between 0 and 1 and iii) indicates the hopping between neighboring sites when it is > 1,. The calculated values of *n* are shown in Fig. 9(b) and found to be less than 1 that recognized the conduction mechanism in our compositions is from the hopping of charge carries at the octahedral sites.

*3.4.3 Dielectric constant*

The dielectric constant provides the information regarding the relative speed of the electromagnetic signal travels in the material. Dielectric properties are influenced by several factors, such as synthesis method, sintering time and temperature, particle size, type and concentration of substituent atom, cation distribution among different sites, etc. When the electromagnetic signal enters into a dielectric material, the microwave speed decreases approximately equal to the $\sqrt{\varepsilon'}$ [46]. The incident wave absorption of the microwave absorber materials can be enhanced and modified by using the complex permittivity. The dielectric constant i.e., real parts of complex permittivity ($\varepsilon'$) represent the storage of electric energy whether imaginary part ($\varepsilon''$) signifies the loss capability. At room temperature, the frequency dependent real part of complex permittivity of the compositions is shown in Fig. 10(a). The dielectric constant decreases with frequency which is the normal behavior of ferrites that is widely studied by many researchers [1, 47]. This dielectric dispersion can be explained based on the Maxwell-Wagner theory of the interfacial polarization **[**48, 49**]** in agreement with the Koop's phenomenological theory **[**50**]**. The theory states that the dielectric medium entails of well conducting grains where they are isolated by the low conducting grain boundaries. The charge carrier are easily transferred the grains and gathered at the grain boundaries under the external applied electric field since the grains conductivity is relatively higher than that of grain boundaries. A large polarization and high dielectric constant produces at low frequency region due to the accumulation of charge carriers at the grain boundaries. The contribution of grain boundaries conductivity at this region is very small. Moreover, high value of dielectric constant at lower frequency can also be described on account of inhomogeneous dielectric structure and consequently interfacial polarization/space polarization. The value of $\varepsilon'$ becomes slower in the mid-frequencies region causes the contribution emanates from the orientational polarization. The dielectric constant relates to the combined comportment of electric charge carriers, electrons and holes. At higher frequency region, the dipolar, ionic and electronic polarization contributes to the dielectric constant. The concentration of $Fe^{2+}$ decreases and at higher frequency the interchange between ferrous ($Fe^{2+}$) and ferric ($Fe^{3+}$) ions does not follow the applied field leading to the decrease of dielectric constant. The space charge polarization starts to move hardly at a certain frequency before retreating and does not play any role for polarization leading the dielectric constant saturated.

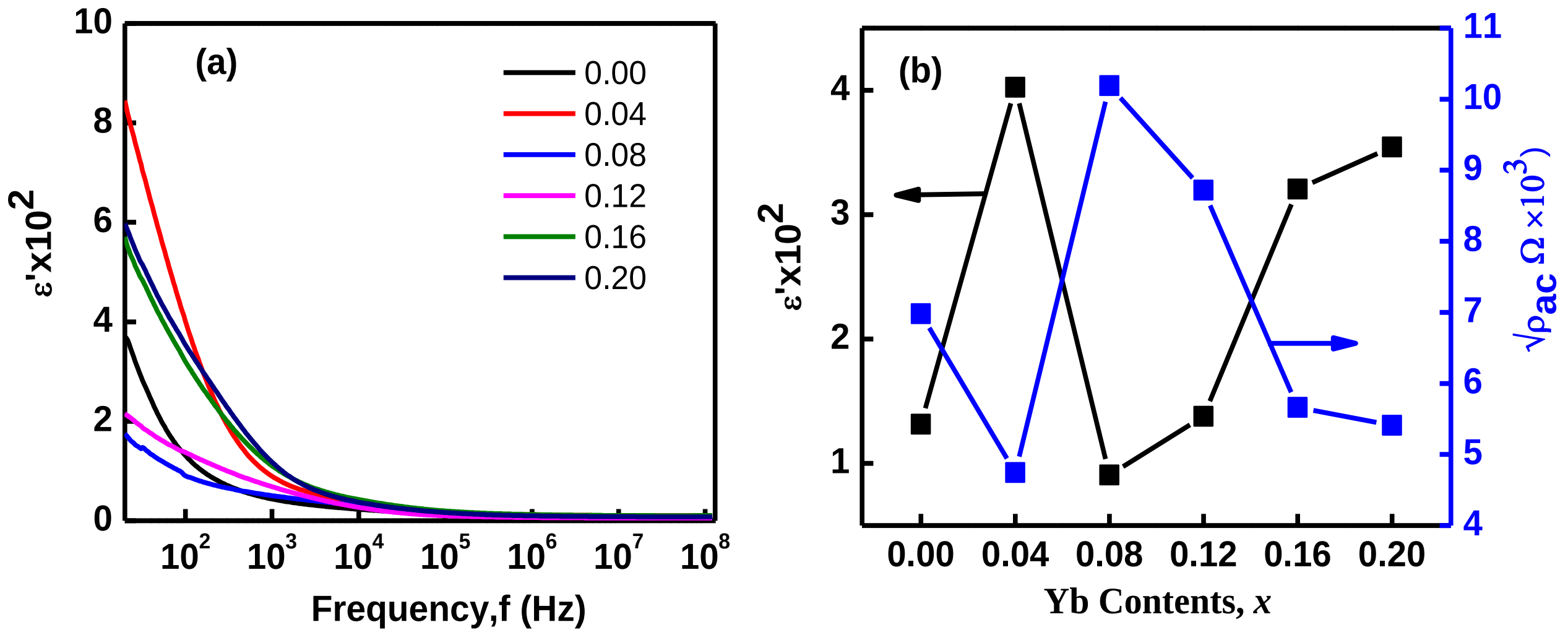

Fig. 10: (a) Frequency dependent dielectric constant and (b) correlation between dielectric constant and AC resistivity with Yb contents of $Ni_{0.5}Zn_{0.5}Yb_xFe_{2-x}O_4$ ($x$= 0.00, 0.04, 0.08, 0.12, 0.16 and 0.20) ferrites at 100 Hz.

There is a correlation between AC electrical resistivity and the dielectric constant. Fig. 10(b) demonstrates the variation of $\sqrt{\rho_{ac}}$ and ε′ which depicts that they are almost inversely proportional to each other. To relate these two parameters the product of ε′ and $\sqrt{\rho_{ac}}$ (at 100 Hz) has been calculated and presented in Table 4. Similar trends have also been reported where the conduction mechanism is controlled by the dielectric polarization [50, 51].

*3. 4. 4 Dielectric loss*

The imaginary part of dielectric dispersion (ε″) of the compositions has been illustrated in Fig. 11(a). It is clear that the value of ε″ reduces fast at low frequency region and becomes slower at high frequency region showing frequency independent behavior similar trends as real part of dielectric constant, ε′ (Fig. 10a). The dielectric loss (*tan*δ) is the loss of electromagnetic radiation in the form of heat due to the collision of atoms in the material during the polarization. The dielectric loss has been determined by the following relation $tan\delta = \varepsilon''/\varepsilon'$ and represented in Fig. 11 (b). The dielectric loss tangent is the angle of dielectric loss and is very important for total core loss. It is clear that the dielectric loss reveals regular Debye relaxation peaks for all compositions. The peak position of the loss tangent shifted towards higher frequency with the increase in $Yb^{3+}$ contents which indicates the increases of jumping probability. The origin of these relaxation peaks can be explicated by the Rezlescu model **[**52**]**.It tells that the influence of n-type (p-type) carriers in the dielectric loss decline slowly (quickly) as the angular frequency upsurges. The influence of the carriers yields a relaxation peak where the frequency of the external applied field exactly matches with the localized electric charge carrier's jumping frequency [53].

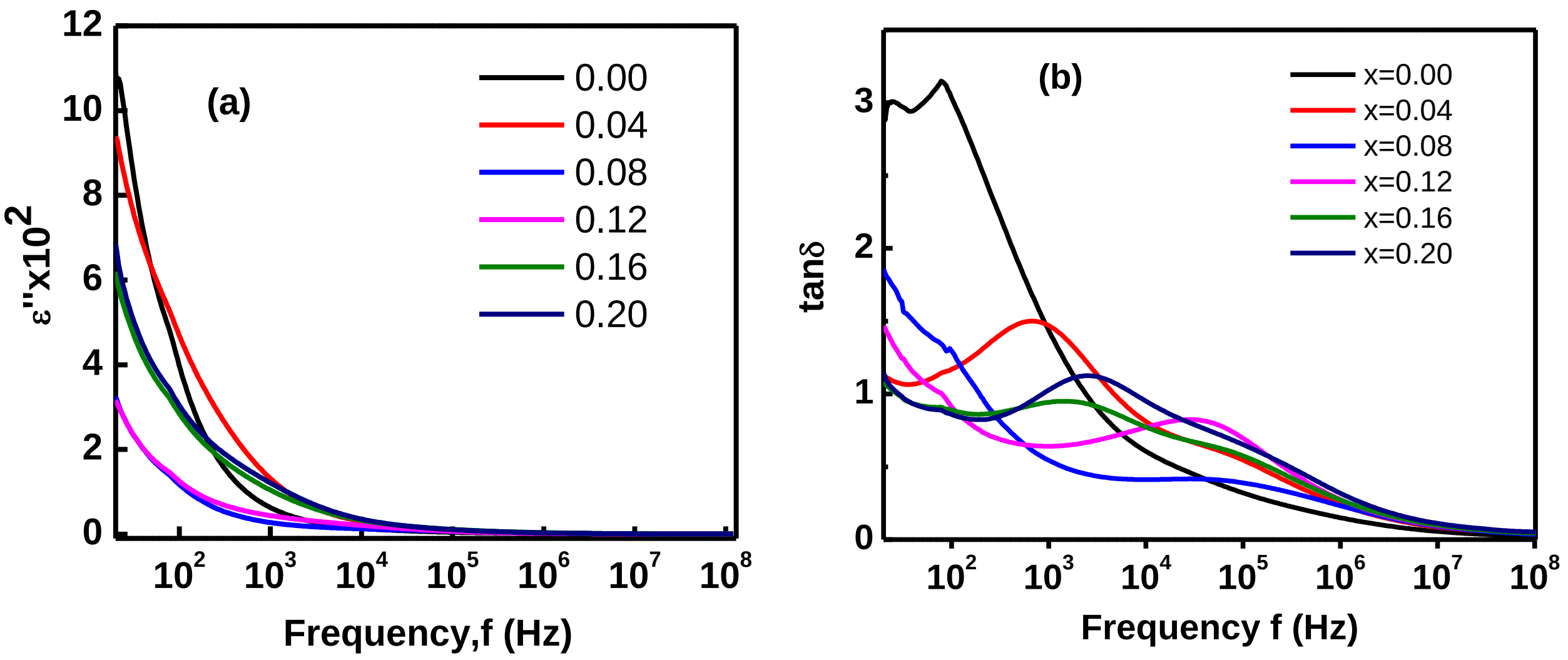

Fig. 11: (a) The variation of imaginary part of dielectric constant as function of frequency (b) dielectric loss with frequency of $Ni_{0.5}Zn_{0.5}Yb_xFe_{2-x}O_4$ ($x$= 0.00, 0.04, 0.08, 0.12, 0.16 and 0.20) ferrites measured at room temperature.

The electron and hole exchange between $Fe^{3+}$, $Fe^{2+}$ and $Ni^{3+}$ and $Ni^{2+}$, respectively contributes to the electric conduction in the compositions. After a certain frequency (15 MHz), the dielectric parameters do not obey the external applied electric field and the polarization cannot match with the field consequently, the dielectric parameters become nearly constant at higher frequency.

Since the value of the dielectric parameters is low, these materials are suggested for high frequency application such as microwave devices.

Table 4: Variation of dielectric parameters with Yb concentration.

| Yb contents (x) | 100 Hz | | | $\sqrt{\rho_{ac}}$(×10[3]) (Ω-cm) | $\varepsilon'\sqrt{\rho_{ac}}$(×10[5]) (Ω-cm) | $\tau_{M''}$ (μsec) |
|---|---|---|---|---|---|---|
| | $\rho_{ac}$(×10[7]) (Ω-cm) | tanδ | ε′ | | | |
| 0.00 | 4.87 | 2.47 | 131.5 | 6.98 | 9.18 | 4.71 |
| 0.04 | 2.25 | 1.16 | 402.3 | 4.75 | 19.09 | 0.91 |
| 0.08 | 10.4 | 1.12 | 90.7 | 10.20 | 9.24 | 0.53 |
| 0.12 | 7.60 | 0.86 | 137.8 | 8.72 | 12.01 | 0.84 |
| 0.16 | 3.20 | 0.85 | 320.5 | 5.66 | 18.15 | 0.69 |
| 0.20 | 2.93 | 0.83 | 354.5 | 5.41 | 19.17 | 0.49 |

*3.4.5 Complex electric modulus spectra*

Complex electric modulus conveys the information regarding the electrical response of the materials that whether polycrystalline samples are homogeneous or inhomogeneous in nature. Frequency dependent real (M′) and imaginary (M″) part of complex electric modulus at room temperature is illustrated in Fig. 12. The complex electric modulus can be expressed by $M^*= M' + iM'' = \varepsilon'/(\varepsilon'^2 + \varepsilon''^2) + i\varepsilon''/(\varepsilon'^2 + \varepsilon''^2)$ [54]. It is seen from the Fig. 12(a), at lower frequency region, the value of M′ shows small value indicating the comfort of polaron hopping. The value of M′ reveals a dispersive maximum in the higher frequency region (not show full spectra due to frequency limitation of our machine).

In Fig. 12(b), the imaginary part of electric modulus (M″) demonstrations that the curves shifts at higher frequency region considerably with significant broadening with increasing Yb contents as well. The peaks shift and broadening at higher frequency are related to the correlation between ion charges and the spread of relaxation time, respectively. The relaxation time ($\tau_{M''}$) has been calculated from the peak position of frequency dependent M″ curve (Fig. 12b) and presented in Table 3. The relaxation time is connected with the jumping probability per unit time P by the relation, $\tau = 1/P$ [55]. In the lower region ($f < f_m$) of maximum peak frequency, the charge carriers are mobile and move over long distances accompanying the hopping conduction process.

However, in the higher region ($f > f_m$), the charge carriers are restricted to potential wells and are movable over short distances accompanying the relaxation polarization process. So the appearance of maximum peak value in the electric modulus indicates the conductivity relaxation (transition from long range to short range mobility with rise of frequency) [54].

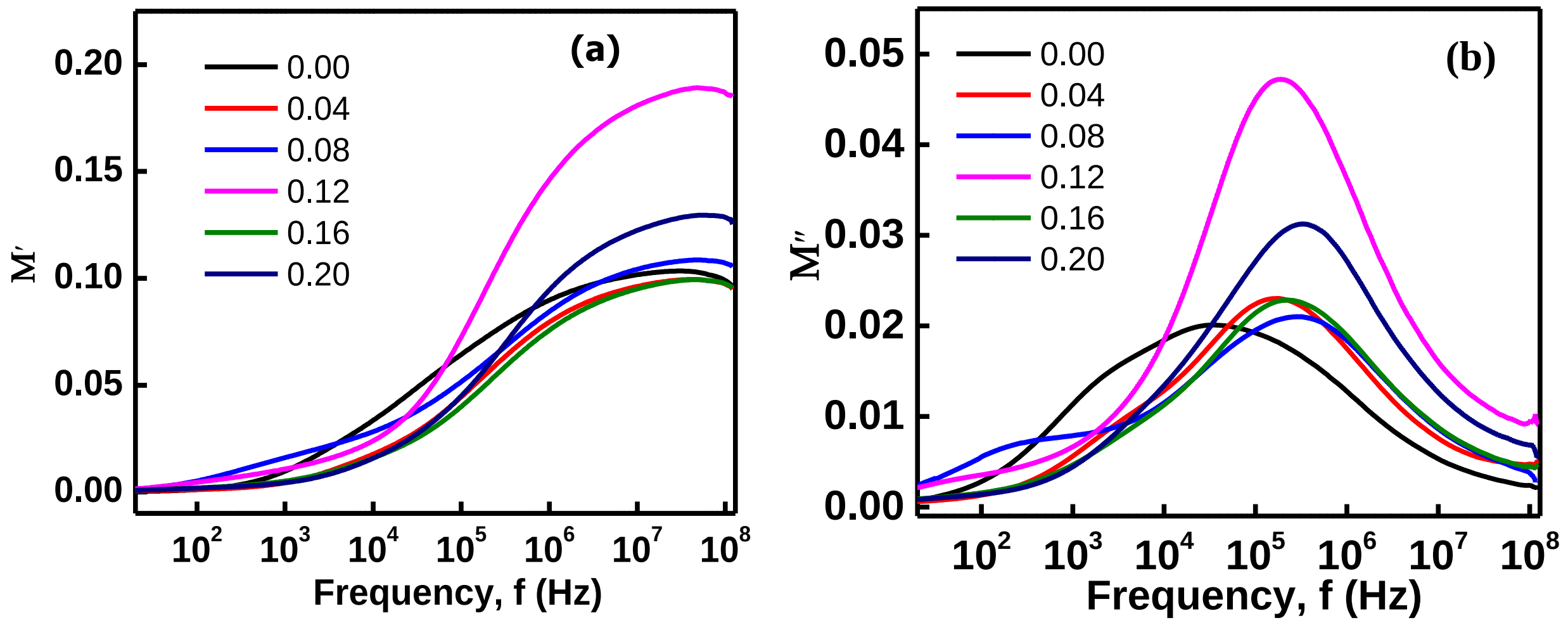

Fig. 12: Frequency dependent real (a) and imaginary (b) part of electric modulus of $Ni_{0.5}Zn_{0.5}Yb_xFe_{2-x}O_4$ ($x$= 0.00, 0.04, 0.08, 0.12, 0.16 and 0.20) ferrites measured at room temperature.

### *3.4.6 Cole-Cole Plot*

The complex electric modulus (Nyquist/Cole-Cole) plot of $Ni_{0.5}Zn_{0.5}Yb_xFe_{2-x}O_4$ ($x$=0.00, 0.04, 0.08, 0.12, 0.16 and 0.20) ferrites sintered at 700 °C has been illustrated in Fig. 13. This plot provides understanding regarding electrical properties, i.e., role of grain and grain boundaries in the electric modulus. There are two semicircles have been perceived in the Cole-Cole plot where the first semicircle characterizes the grain boundary contribution at low frequency region and the second semicircle denotes the grain involvement at high frequency of the materials. Depending on the strength of relaxation and the available frequency range, the Cole-Cole plot show full, partial or absent of any semicircle. It is seen from Fig. 13, two semicircles are partially overlapped that their centers are lying under the real axis indicating the non-Debye type of relaxation process. Inset shows the first semicircle at lower frequency region. The small semicircle's attributes to the grain boundary effect and large semicircle is assumed to be persuaded by the grain effect which arises from smaller capacitance dominating the electric modulus. Two semicircles in the plot suggest significant influences of grain and grain boundary contribution in the studied compositions. It is bit difficult to observe two complete semicircles due to huge resistance alteration between grain and grain boundary since the electric modulus represents the smallest capacitance while the impedance plot demonstrates the largest resistance.

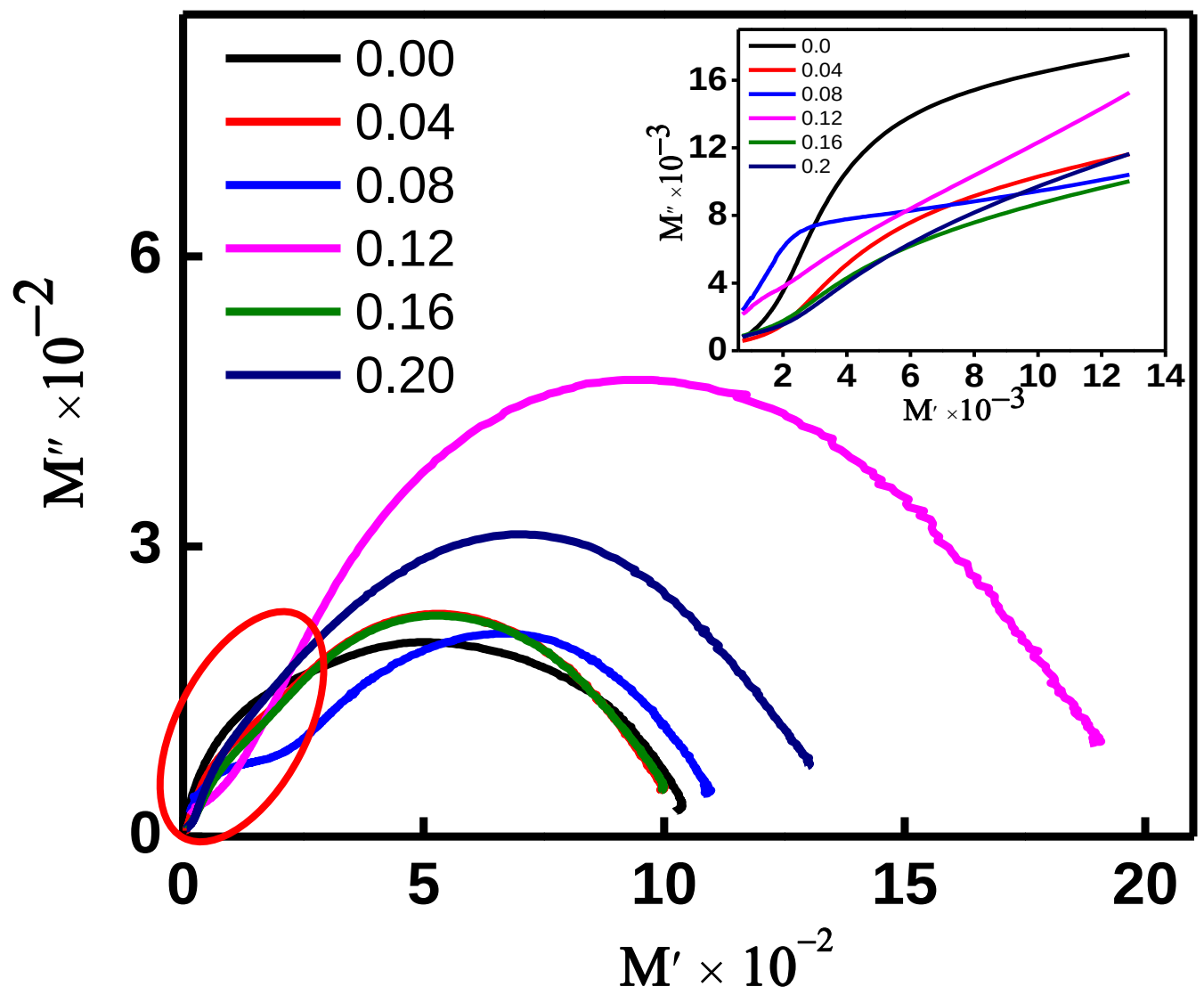

Fig. 13: Room temperature Nyquist/Cole-Cole plot of electric modulus of $Ni_{0.5}Zn_{0.5}Yb_xFe_{2-x}O_4$ ($x$= 0.00, 0.04, 0.08, 0.12, 0.16 and 0.20) ferrites sintered at 700 °C. Inset: magnification of low frequency semi-circle (red circle) for better understanding.

*3.5. Diffuse reflectance spectroscopy (DRS)*

The optical band gap of all the compositions of $Ni_{0.5}Zn_{0.5}Yb_xFe_{2-x}O_4$ has been measured from the Tauc's plot using UV-Vis diffuse reflectance spectroscopy shown in Fig. 14. The Tauc's relation is given as [56], $F(R_\infty) = \frac{A(h\nu - E_g)^n}{h\nu}$, where, $F(R_\infty) = \frac{(1-R_\infty)^2}{h\nu}$ is the Kubelka-Munk function, $R_\infty$ is the ratio of diffuse reflectance between the sample and the reference material, A is a constant, hν is the incident photon energy. A graph has been plotted of $[F(R_\infty)h\nu]^2$ against hν (Fig. 14). From the extrapolating of the linear part of the plot to the energy axis (intercepting value) provides the value of the energy band gap ($E_g$). The value of $E_g$ is found to be 2.73 eV to 3.25 eV. Our calculated parent composition ($Ni_{0.5}Zn_{0.5}Fe_2O_4$) $E_g$ value (2.73 eV) is in good agreement with the reported value of 2.56 eV **[**57**]**. It shows that the $E_g$ value increases with the increase in the Yb content. It is suggested that the value of $E_g$ can be changed by different factors such as crystallite size, structural parameter, and presence of impurities. The increase of $E_g$ in this case may be ascribed to the smaller crystallite size with increasing Yb contents in the composition.

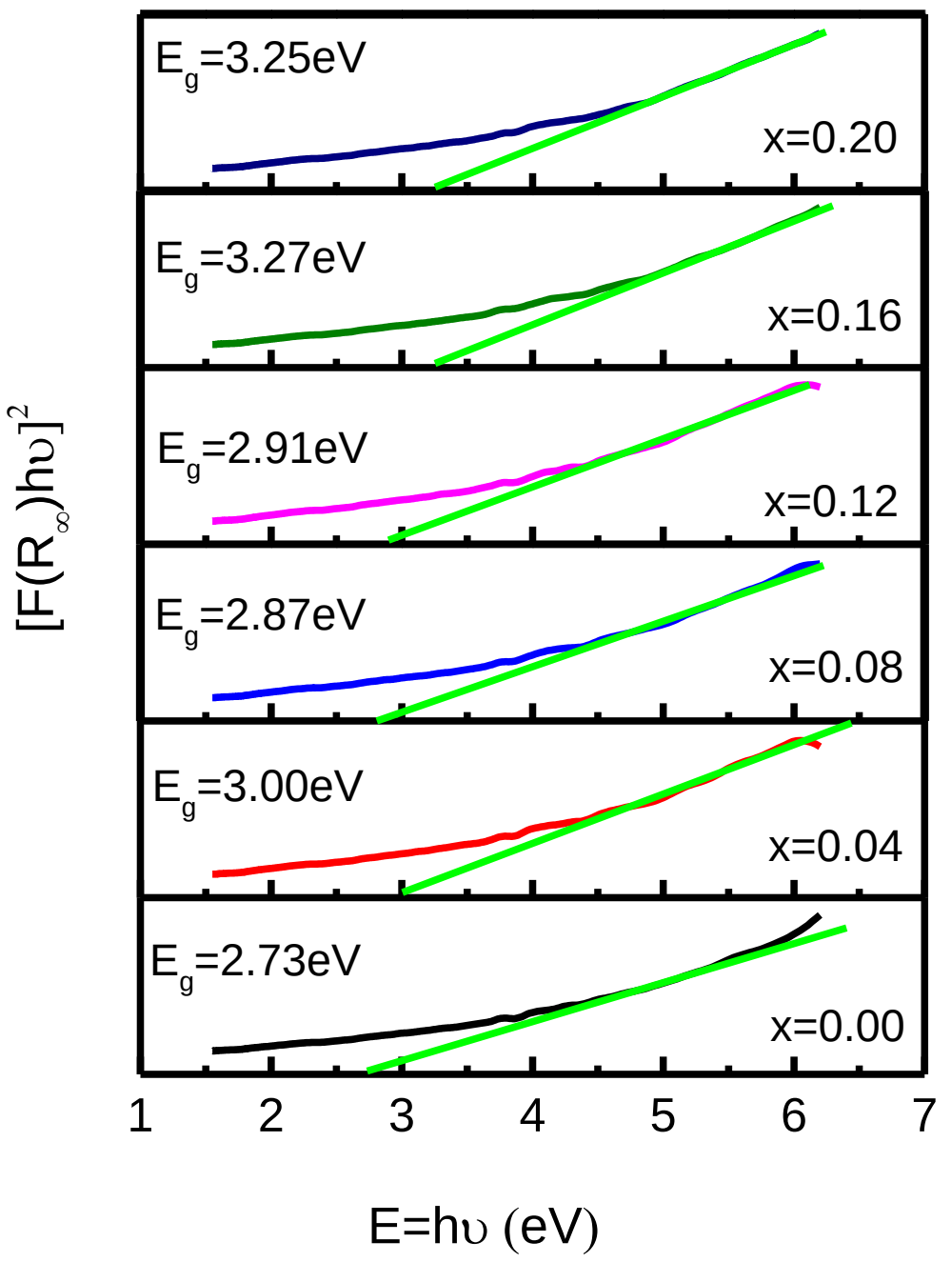

Fig. 14: The UV-Vis spectra of $Ni_{0.5}Zn_{0.5}Yb_xFe_{2-x}O_4$ ($x$= 0.00, 0.04, 0.08, 0.12, 0.16 and 0.20) sintered at 700 °C.

## *4. Conclusions*

The compositions $Ni_{0.5}Zn_{0.5}Yb_xFe_{2-x}O_4$ ($x$= 0.00, 0.04, 0.08, 0.12, 0.16 and 0.20) sintered at 700 °C have been successfully synthesized by sol-gel auto combustion technique. The particle size, average crystallite and grain size have been estimated in the range of 183 to 343 nm , 64 to 11 nm and 52 to 18 nm for different $Yb^{3+}$content by the dynamic light scattering (DLS) technique, XRD spectra and FESEM images, respectively and correlation between them is successfully explained. Two expected vibration bands are found to be at ~595 $cm^{-1}$ corresponds to M–O bond in the *A*-sites and the other at ~360 $cm^{-1}$ corresponds to M–O bond in the *B*-sites by the FTIR spectra that ratify completion of successful sample synthesis. Temperature dependent resistivity explores the semiconducting nature of the samples. Frequency dependent dielectric dispersion has been observed and explained using the Maxwell-Wagner theory of the interfacial polarization in settlement with the Koop's phenomenological theory on account of polarization due to mobility of charge between grain and grain boundary. The dielectric loss reveals regular Debye relaxation peaks for all compositions that have successfully been explained by Rezlescu model considering combined impact of n-type and p-type charge carriers. The estimated relaxation time is found to be in the range of 4 to 0.5 micro seconds. Energy band gap escalates (2.73 eV to 3.25 eV) with rising substituted Yb contents that is attributed from the UV-Vis spectroscopy. It is remarkable that the estimated high dc resistivity value is found to be in the range of $2.2 \times 10^9$ Ω-cm to $2.6 \times 10^{10}$ Ω-cm for different $Yb^{3+}$ substitution (*x*= 0.0 to *x*= 0.20 in step of 0.04) is the silent feature of these ferrites, thereby lowering dielectric loss makes them suitable candidate for implication in high frequency applications such as microwave devices.

## Acknowledgments

The authors are grateful to the Directorate of Research and Extension (DRE), Chittagong University of Engineering and Technology, Chattogram-4349, Bangladesh under grant number CUET DRE (CUET/DRE/2016-2017/PHY/003) for arranging the financial support for this work. We are thankful for the laboratory support of the Materials Science Division, Atomic Energy Center, Dhaka 1000, Bangladesh.